\documentstyle[12pt,epsf]{article}
\newcommand{\bfig}{\begin{figure}}
\newcommand{\efig}{\end{figure}}

\def\one-loop{\mbox{\scriptsize one-loop}}
\def\cD{{\cal D}}
\def\a{\alpha}

\def\l{\lambda}
\def\lb{\bar{\lambda}}
\def\s{\sigma}
\def\th{\theta}
\def\thb{\bar{\theta}}

\def\psivector{\mbox{\boldmath$\psi_{Q}$}}
\def\Upsilonvector{\mbox{\boldmath$\Upsilon$}}
\def\Qvector{\mbox{\boldmath$Q$}}
\def\Omegavector{\mbox{\boldmath$\Omega$}}
\def\Pivector{\mbox{\boldmath$\Pi$}}
\def\Zvector{\mbox{\boldmath$Z$}}
\def\Db{\bar{D}}
\def\dg{\dagger}
\def\half{\frac{1}{2}}
\def\ud{\underline}

\def\G{\Gamma}

\catcode`\@=11

\@addtoreset{equation}{section}
\def\theequation{\arabic{section}.\arabic{equation}}

\def\@normalsize{\@setsize\normalsize{15pt}\xiipt\@xiipt
\abovedisplayskip 14pt plus3pt minus3pt%
\belowdisplayskip \abovedisplayskip
\abovedisplayshortskip  \z@ plus3pt%
\belowdisplayshortskip  7pt plus3.5pt minus0pt}

\def\small{\@setsize\small{13.6pt}\xipt\@xipt
\abovedisplayskip 13pt plus3pt minus3pt%
\belowdisplayskip \abovedisplayskip
\abovedisplayshortskip  \z@ plus3pt%
\belowdisplayshortskip  7pt plus3.5pt minus0pt
\def\@listi{\parsep 4.5pt plus 2pt minus 1pt
            \itemsep \parsep
            \topsep 9pt plus 3pt minus 3pt}}

\def\underline#1{\relax\ifmmode\@@underline#1\else
        $\@@underline{\hbox{#1}}$\relax\fi}
\@twosidetrue

\relax

\catcode`@=12

\catcode`\@=11

\def\section{\@startsection{section}{1}{\z@}{3.5ex plus 1ex minus
   .2ex}{2.3ex plus .2ex}{\large\bf}}

\def\thesection{\Roman{section}.}

\def\appendix{\setcounter{section}{0}
        \def\thesection{Appendix }
        \def\theequation{\Alph{section}.\arabic{equation}}}

\def\ps@headings{\def\@oddfoot{}\def\@evenfoot{}
\def\@oddhead{\hbox{}\hfill
        \makebox[.5\textwidth]{\raggedright\ignorespaces --\thepage{}--
        \hfill {}}}
\def\@oddhead{\hbox{}\hfill --\thepage{}-- \hfill
        {}}
\def\@evenhead{\@oddhead}
\def\subsectionmark##1{\markboth{##1}{}}
}

\ps@headings

\catcode`\@=12

\relax

\def\figcap{\section*{Figure Captions\markboth
        {FIGURECAPTIONS}{FIGURECAPTIONS}}\list
        {Fig. \arabic{enumi}:\hfill}{\settowidth\labelwidth{Fig. 999:}
        \leftmargin\labelwidth
        \advance\leftmargin\labelsep\usecounter{enumi}}}
 \relax
\def\tablecap{\section*{Table Captions\markboth
        {TABLECAPTIONS}{TABLECAPTIONS}}\list
        {Table \arabic{enumi}:\hfill}{\settowidth\labelwidth{Table 999:}
        \leftmargin\labelwidth
        \advance\leftmargin\labelsep\usecounter{enumi}}}
 \relax
\def\reflist{\section*{References\markboth
        {REFLIST}{REFLIST}}\list
        {[\arabic{enumi}]\hfill}{\settowidth\labelwidth{[999]}
        \leftmargin\labelwidth
        \advance\leftmargin\labelsep\usecounter{enumi}}}
 \relax

\catcode`\@=11

\def\ps@headings{\def\@oddfoot{}\def\@evenfoot{}
\def\@oddhead{\hbox{}\hfill
        \makebox[.5\textwidth]{\raggedright\ignorespaces --\thepage{}--
        \hfill {}}}
\def\@evenhead{\@oddhead}
\def\subsectionmark##1{\markboth{##1}{}}
}

\ps@headings

\relax

\newskip\humongous \humongous=0pt plus 1000pt minus 1000pt

\newif\ifdtup

\def\beq{\begin{equation}}
\def\eeq{\end{equation}}

\def\beqn{\begin{eqnarray}}
\def\eeqn{\end{eqnarray}}
\relax

\def\G2{{\; \rm GeV/}c^2}
\def\G{\; \rm GeV}

\def\dotx{\dotx{\dot\overline{x}}}

\relax

\def\p{\partial}

\begin{document}
\begin{titlepage}
\begin{flushright}
       {\normalsize OU-HET 306 \\  hep-th/9812177\\
           December, 1998 }
\end{flushright}
%
\begin{center}
  {\large \bf $USp(2k)$ Matrix Model:  \\
     Schwinger-Dyson Equations and  \\
 Closed-Open String Interactions}
\footnote{This work is supported in part
 by the Grant-in-Aid  for Scientific Research (10640268,10740121)
 and Grant-in-Aid  for Scientific Research fund (97319)
from the Ministry of Education, Science and Culture, Japan.}

\vfill
         {\bf H.~Itoyama}  \\
            and \\
              {\bf A.~Tsuchiya}\\
        Department of Physics,\\
        Graduate School of Science, Osaka University,\\
        Toyonaka, Osaka 560-0043, Japan\\
\end{center}
\vfill
\begin{abstract}
 We derive the Schwinger-Dyson/loop equations for the $USp(2k)$ matrix model
  which  close among the closed and open Wilson loop variables. These loop
 equations exhibit a complete set of the joining and splitting interactions
  required for the nonorientable $TypeI$ superstrings.  The open loops
    realize the $SO(2n_{f})$ Chan-Paton factor and  their
 linearized loop equations derive the mixed Dirichlet/Neumann
 boundary conditions.
 
\end{abstract}
\vfill
\end{titlepage}

\section{Introduction}
 Intensive study has been recently made for the nonperturbative
 formulation of superstrings, trying to uncover the properties
 which are not accessible in the first quantized 
 perturbative formulation of superstrings. The study in this direction
 appears to be imperative in order to confront string physics with the world
 we observe in nature.
  One such approach toward this goal starts with a model as
 the constructive definition of the type $IIB$ or the type $I$ superstrings
 in the form of zero dimensional  reduced model \cite{IKKT,FKKT,IT1,IT2} and 
 this direction, in particular, the  case of  the $USp$ matrix model
 \cite{IT1,IT2} for the type $I$ superstrings is the focus of the present
 paper.

  The reduced model is a nonabelian counterpart of the first quantized
  critical superstring theory in the Schild gauge \cite{Schild} and the
 large $k$ limit makes this connection clear.  From this point alone, it is
 certain that the model is for unification of all forces including gravity
 and is not limited to low energy phenomena mediated by open strings. Another
 aspect of the model is that the matrix degrees of freedom in fact generate
 manybody effects of strings albeit the fact that the model is originally
 in its first quantized form.  We would appreciate these points better if we
 are able to formulate the model in the second quantized form.
  The Schwinger-Dyson equations representing loop dynamics accomplish this.

   The Schwinger-Dyson equations for the case of the $IIB$  matrix model
  have already been examined in \cite{FKKT}. 
Here we focus on  the derivation of the Schwinger-Dyson equations
   for the case of  the $USp$  matrix model.  The properties  and implications
  of the $USp$  matrix model have been elaborated in \cite{IT1,IT2,IM,CIK}.
   Among other things, this model introduces  open string degrees of freedom
 in an explicit way  in contrast to the ones associated with
  D-objects \cite{Pol} as classical solutions.  This point translates into
  the open loop variables of our paper.
  In the next section, we  specify a set of loop variables   adopted 
 for the Schwinger-Dyson equations of the $USp(2k)$  matrix 
  model.  The $SO(2n_{(f)})$  Chan-Paton factor emerging from the open loop is
   observed.
  In section  three, we  derive the Schwinger-Dyson equations
   and  a complete set of the joining and splitting interactions
  required for the nonorientable $TypeI$ superstrings  is exhibited.
   Comparison with the string field theory of \cite{GSSFT,KT} is made.
  In section four, we study these equations at the linearized level.
  In addition to the Virasoro condition  for the closed loops noted
 before at \cite{FKKT}, we find that the open loops satisfy
  the appropriate mixed Dirichlet/Neumann boundary conditions.
   The final section is devoted to outlook and open questions  for
  the reduced model in general.

In Appendix A, we  present the action of the $USp$ matrix model
in a more compact component form  than is presented in \cite{IT1, IT2}, so
 that the derivation in section three becomes a more
 manageable procedure.  Readers are advised to look at some of the notation
  establised here before going into the text.
  In Appendix B,  we list kinetic terms of the Schwinger-Dyson equations.

\section{Choice of variables}
  Let us first introduce a discretized path-ordered exponential
 which represents a configuration of a string in momentum superspace:
\beqn
U[p^{M}_{.},\eta_{.};n_1,n_0] \equiv
P \exp (-i \sum_{n=n_0}^{n_1} (p^{M}_{n} v_{M}+\bar{\eta}_{n} \Psi))
= \stackrel{\leftarrow}{\prod_{n=n_0}^{n_1}} 
 \exp (-i p^{M}_{n} v_{M}-i \bar{\eta}_{n} \Psi)\;\;,
\eeqn
  where  $p^{M}_{n}$ and $\eta_n$ are respectively the sources
 or the momentum distributions for $v_{M}$ and those for $\Psi$.
The closed loop is then defined by
\beq
\Phi[p^{M}_{.},\eta_{.};n_1,n_0] \equiv Tr U[p^{M}_{.},\eta_{.};n_1,n_0]\;\;.
\eeq
To consider an open loop, let us introduce
$\Xi = \left( \xi, \xi^{*} \right)$ as bosonic sources for $\Qvector_{(f)}$
 and $\Qvector_{(f)}^{\ast}$, and $\Theta = \left(\theta, \bar{\theta} \right)$
as  Grassmannian ones for $\psivector_{(f)}$ and $\psivector_{(f)}^{\ast}$:
$ \left( \Xi  \Omegavector_{(f)} \right)  = \xi \Qvector_{(f)}
                      +F^{-1} \xi^{\ast}\Qvector_{(f)}^{\ast}$,
 $\left( \Theta \Upsilonvector_{(f)} \right) = \theta \psivector_{(f)} 
+F^{-1} \bar{\theta} \psivector^{\ast}_{(f)}.$
   We write these collectively as
\beq
\left( \Lambda \Pivector_{(f)} \right) \equiv \left( \Xi \Omegavector_{(f)}
 \right)
 +
                      \left( \Theta  \Upsilonvector_{(f)}  \right)\;\;.
\eeq
  The open loop is defined by
\beq
\Psi_{f'f}[k^{m}_{.},\zeta;l_1,l_0;\Lambda',\Lambda]
\equiv   \left( \Lambda'  \Pivector_{(f')} \right)
F U[k^{m}_{.},\zeta_{.};l_1,l_0] 
\left( \Lambda  \Pivector_{(f)} \right) \;\;,
\eeq
  where $f$ and $f'$  are the Chan-Paton indices.
 In view of the notion of  the macroscopic loop in the one and multi matrix
 models  of random surfaces, it is clear that these loops are the appropriate
nonabelian generalization to the reduced model for string
unification and that they generate  all of the observables in the theory
under question.
 
  Let us see how the nonorientability of the closed and the open
 strings  is realized in the loops  we introduced. 
Using the eqs. (\ref{eq:adj}), (\ref{eq:asym}), namely,
$v^{t}_{M}=\mp F v_{M} F^{-1}, \Psi^{t}= \mp F \Psi F^{-1},$
 and $F^{t}= -F$ , we  readily obtain
\beqn
\Phi[p^{M}_{.},\eta_{.};n_1,n_0] 
=Tr(\stackrel{\rightarrow}{\prod_{n=n_0}^{n_1}} 
\exp (-i p^{M}_{n} v^{t}_{M}-i \bar{\eta}_{n} \Psi^{t}))
=\Phi[\mp p^{M}_{.},\mp \eta_{.};n_0,n_1]\;\;,
\label{csnonori}
\eeqn
 and
\beq
\Psi_{f'f}[k^{M}_{.},\zeta_{.};l_1,l_0;\Lambda',\Lambda]
=-\Psi_{ff'}[\mp k^{M}_{.},\mp \zeta_{.};l_0,l_1;\Lambda,\Lambda']\;\;.
\label{osnonori}
\eeq
These equations relate a string configuration to the one with its
orientation reversed. The Chan-Paton factor is reversed as well for the
 case of the open loops. The minus signs in front of
 $p^{m}_{.}$, $\eta_{.}$, $k^{m}_{.}$ and $\zeta_{.}$ in eq.
 (\ref{csnonori}) and  eq. (\ref{osnonori}) reflect the orientifold
structure of the $USp(2k)$ matrix model.
The overall minus sign in the last line of (\ref{osnonori}) 
  comes from $F^{t}=-F$ of the $usp$ Lie algebra and
 corresponds to the $SO(2 n_f)$ gauge group. ( Clearly we obtain
 the plus sign for the case of the $so$ Lie algebra: $F^{t}= F$.)
  We see that the infrared stability of perturbative vacua
  \cite{GS,IMo} tells
that the original matrices must be based on
the $usp$ as opposed to the $so$ Lie algebra and that $n_f =16$.
This  latter property also follows from the anomaly cancellation
of the T-dualized  representation of the theory  by the
 $6D$ worldvolume gauge theory \cite{IT2}.

\section{Schwinger-Dyson equations}
To proceed to the loop equations, let us first introduce abbreviated notation:
\beqn
\Phi[(i)] &\equiv& \Phi[p^{(i)}_{.},\eta^{(i)}_{.};n^{(i)}_{1},n^{(i)}_{0}]\;,
 \;\Psi[(i)] \equiv
\Psi_{f^{(i)'}f^{(i)}}[k^{(i)}_{.},\zeta_{.}^{(i)};l^{(i)}_{1},l^{(i)}_{0};
\Lambda^{(i)'},\Lambda^{(i)}]\;,  \nonumber\\
\int d\mu  \cdots &\equiv& \int [dv][d\Psi][d\Qvector][d\Qvector^{\ast}]
                        [d\psivector][d\psivector^{\ast}]  \cdots \;.
\nonumber 
\eeqn
We begin with the following set of equations consisting of $N$ closed loops 
and $L$ open loops:
\beqn
&&0=\int d\mu
\frac{\partial}{\partial X^{r}}
\left\{ Tr(U[p^{(1)}_{.},\eta^{(1)}_{.};n^{(1)}_{2},n^{(1)}_{1}+1] T^r 
U[p^{(1)}_{.},\eta^{(1)}_{.};n^{(1)}_{1},n^{(1)}_{0}]) \right. \nonumber\\
&& \hspace*{4.5cm}
\left.  \Phi[(2)] \cdots \Phi[(N)]  \Psi[(1)] \cdots \Psi[(L)] 
\, e^{-S} \right\} \;\;,
\label{loopeq1}
\\
&& 0=\int d\mu \frac{\partial}{\partial X^{r}}
\left\{  \left( \Lambda^{(1)'}  \Pivector_{(f^{(1)'})}  \right) 
F U[k^{(1)}_{.},\zeta^{(1)}_{.};l^{(1)}_{2},l^{(1)}_{1}+1]T^r
U[k^{(1)}_{.},\zeta^{(1)}_{.};l^{(1)}_{1},l^{(1)}_{0}]
 \left( \Lambda^{(1)} \Pivector_{(f^{(1)})}  \right) \right.  \nonumber\\
&&\hspace*{4.5cm}
 \left. \Phi[(1)] \cdots \Phi[(N)] 
\Psi[(2)] \cdots \Psi[(L)]
\, e^{-S} \right\} \;\;,
\label{loopeq2}
\\
&& 0=\int d\mu
\frac{\partial}{\partial \Zvector_{(f) i}}
\left\{ (U[k^{(1)}_{.},\zeta^{(1)}_{.};l^{(1)}_{1},l^{(1)}_{0}]
\left( \Lambda^{(1)}  \Pivector_{(f^{(1)})} \right)  )_{i}  \right. \nonumber\\
&& \hspace*{5cm}  \left. \Phi[(1)] \cdots \Phi[(N)]  \,
\Psi[(2)] \cdots \Psi[(L)]
\, e^{-S} \right\} \;\;,
\label{loopeq3}
\eeqn
where $X^r$  denotes $v_{M}^{r}$ or $\Psi^r_{\alpha}$ while
 $\Zvector_{(f)i}$ denotes
$\Qvector_{(f)i}$ or $\psivector_{(f) i \alpha}$.

\begin{figure}
\centerline{\epsfbox{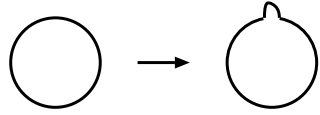}}
\caption{\small infinitesimal deformation of a closed string}
\label{closedkin}
\end{figure}

\begin{figure}
\vspace*{1cm}
\centerline{\epsfbox{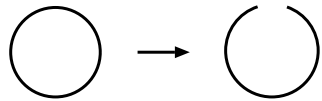}}
\caption{\small closed-open transition}
\label{closed-open}
\end{figure}

\begin{figure}
\vspace*{1cm} 
\centerline{\epsfbox{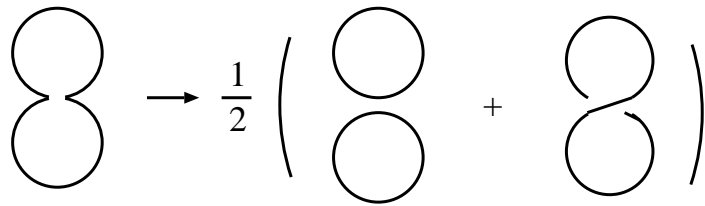}}
\caption{\small splitting and twisting of a closed string}
\label{closed}
\end{figure}

\begin{figure}
\vspace*{1cm} 
\centerline{\epsfbox{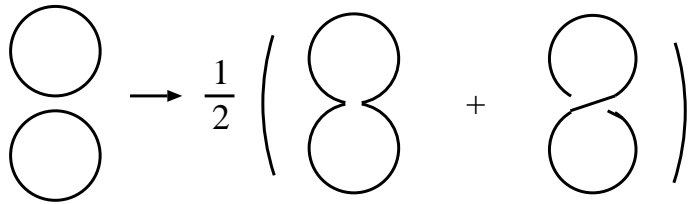}}
\caption{\small joining of two closed strings}
\label{closedclosed}
\end{figure}

\begin{figure}
\vspace*{1cm}
\centerline{\epsfbox{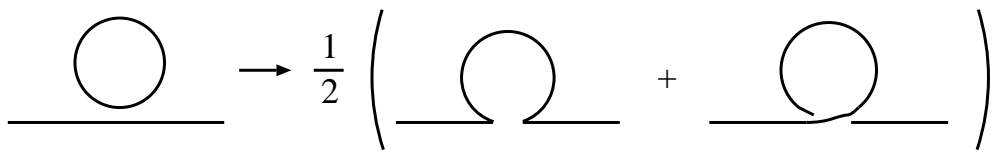}}
\caption{\small joining of a closed string and an open string}
\label{openclosed}
\end{figure}

In what follows, we will exhibit eqs.
(\ref{loopeq1}) $\sim$ (\ref{loopeq3}) in the form of loop equations
(\ref{loopeq1complete}) $\sim$ (\ref{loopeq3complete}).
 We will repeatedly use
\beq
\sum_{r=1}^{2 k^2 \pm k} (T^r)_{i}^{\; j} (T^r)_{k}^{\; l}
=\frac{1}{2} (\delta_{i}^{\; l} \delta^{j}_{\; k}
\mp F^{-1}_{ik} F^{lj}) \;\;,
\eeq
  which is nothing but the expression for the projector (\ref{eq:projector}).
In these equations below, 
\beqn
P^{(i)}_{n}=\left\{ \begin{array}{l} 
                     p^{(i)M}_{n} \; \mbox{if} \; X^r=v^{r}_{M} \;, \\
                     -\bar{\eta}^{(i)}_{n}  \; \mbox{if} \; X^r=\Psi^r \;,
                    \end{array} \right.   \;\;\;
K^{(i)}_{n}=\left\{ \begin{array}{l} 
                     k^{(i)M}_{n} \; \mbox{if} \; X^r=v^{r}_{M} \;, \\
                     -\bar{\zeta}^{(i)}_{n}  \; \mbox{if} \; X^r=\Psi^r \;,
                    \end{array} \right. 
\eeqn               
and $\Lambda^{(i)}$ not multiplied by $\Pi$  represents either
 $\Xi^{(i)}$ or $\Theta^{(i)}$.
The symbol $\hat{b}$  denotes an omission of 
 the $b$-th closed or open loop.

\beqn
\bullet
(\ref{loopeq1}) &\Rightarrow&  \; 0  =
(1)\; \mbox{\underline{kinetic term (Fig. \ref{closedkin}),
 \ref{closed-open})}}
\;+\; (2)\; \mbox{\underline{splitting and twisting
 (Fig. \ref{closed})}} \\
&+& (3)\; \mbox{\underline{joining with a closed string
 (Fig. \ref{closedclosed})}}
\;+\; (4)\; \mbox{\underline{joining with an open string
 (Fig. \ref{openclosed})}} \;\;.  \nonumber
\eeqn
Here
\beqn
 &(1)& = 
\frac{1}{g^2} \left\langle  \left( \delta_{X} \Phi[(1); X^r]  \right) 
\Phi[(2)] \cdots \Phi[(N)]  \Psi[(1)] \cdots \Psi[(L)] \right\rangle \;\;,
  \label{eq:loop1kin} \\
&(2)& =
\left\langle
\left( -\frac{i}{2}\sum_{n=n^{(1)}_{0}}^{n^{(1)}_{1}} P^{(1)}_{n}  \right)
 \left\{ \Phi[p^{(1)}_{.},\eta^{(1)}_{.};n^{(1)}_{1},n+1]
\Phi[p^{(1)}_{.},\eta^{(1)}_{.};n,n^{(1)}+1] \right.\right.\nonumber\\
 &\pm &   \left. Tr(U[p^{(1)}_{.},\eta^{(1)}_{.};n,n^{(1)}_1+1]
U[\mp p^{(1)}_{.},\mp \eta^{(1)}_{.};n+1,n^{(1)}_1]) \right\}
 \left. 
\Phi[(2)] \cdots \Phi[(N)]  \Psi[(1)] \cdots \Psi[(L)] \right\rangle
\nonumber\\
&+& \left\langle
 \left( -\frac{i}{2}\sum_{n=n^{(1)}_{1}+1}^{n^{(1)}_{2}} P^{(1)}_{n} \right)
 \left\{ \Phi[p^{(1)}_{.},\eta^{(1)}_{.};n,n^{(1)}_{1}+1]
\Phi[p^{(1)}_{.},\eta^{(1)}_{.};n^{(1)},n+1] \right.\right.\nonumber\\
 &\pm& \!\!\!\!\left. Tr(U[p^{(1)}_{.},\eta^{(1)}_{.};n^{(1)}_{1},n+1]
U[\mp p^{(1)}_{.},\mp \eta^{(1)}_{.};n^{(1)}_{1}+1,n]) \right\}
  \left. 
\Phi[(2)] \cdots \Phi[(N)]  \Psi[(1)] \cdots \Psi[(L)] \right\rangle \;\;, \\
 &(3)& = 
\left\langle
\left( -\frac{i}{2}\sum_{b=2}^{N} \sum_{n=n^{(b)}_{0}}^{n^{(b)}_{1}}
 P^{(b)}_{n} \right)
 \left\{ Tr(U[p^{(1)}_{.},\eta^{(1)}_{.};n^{(1)}_{1},
n^{(1)}_{1}+1] U[p^{(b)}_{.},\eta^{(b)}_{.};n,n+1]) \right.\right.\nonumber\\
&& \mp \left. Tr(U[\mp p^{(1)}_{.},\mp
 \eta^{(1)}_{.};n^{(1)}_{1}+1,n^{(1)}_{1}]
U[p^{(b)}_{.},\eta^{(b)}_{.};n,n+1]) \right\}  \nonumber\\
&&\left. \Phi[(2)] \cdots \hat{b} \cdots \Phi[(N)]  
\Psi[(1)] \cdots \Psi[(L)] \right\rangle \;\;, \\
 &(4)& =
\left\langle 
\left( -\frac{i}{2}\sum_{b=1}^{L} \sum_{l=l^{(b)}_{0}}^{l^{(b)}_{1}}
 K^{(b)}_{l} \right) \right. \nonumber \\
  && \left\{  \left( \Lambda^{(b)'} \Pivector_{(f^{(b)'})} \right) F
U[k^{(b)}_{.},\zeta^{(b)}_{.};l^{(b)}_{1},l+1]
U[p^{(1)}_{.},\eta^{(1)}_{.};n^{(1)}_{1},n^{(1)}_{1}+1]
U[k^{(b)}_{.}, \zeta^{(b)}_{.};l,l^{(b)}_{0}]
 \left(  \Lambda^{(b)}  \Pivector_{(f^{(b)})} \right) \right. \nonumber\\
 &\mp& \!\!\!\!\left. \left( \Lambda^{(b)'}  \Pivector_{(f^{(b)'})}  \right) F
U[k^{(b)}_{.},\zeta^{(b)}_{.};l^{(b)}_{1},l+1]
U[\mp p^{(1)}_{.},\mp \eta^{(1)}_{.};n^{(1)}_{1}+1,n^{(1)}_{1}]
U[k^{(b)}_{.},\zeta^{(b)}_{.};l,l^{(b)}_{0}]
\left(  \Lambda^{(b)} \Pivector_{(f^{(b)})} \right) \right\} \nonumber\\
&& \hspace*{4cm}\left. 
\Phi[(2)] \cdots \Phi[(N)]  
\Psi[(1)] \cdots \hat{b} \cdots \Psi[(L)] \right\rangle \;\;\;.
\label{loopeq1complete}
\eeqn
  We spell out   the explicit form of  $\delta_{X} \Phi[(1); X^r]$
 in Appendix B.   ( This term comes from the variation of the action and
 contains terms representing closed-open transition.)

 

\begin{figure}
\vspace*{1cm} 
\centerline{\epsfbox{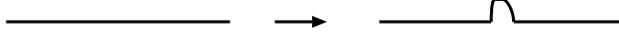}}
\caption{\small infinitesimal deformation of an open string: case one}
\label{openkin1}
\end{figure}

\begin{figure}
\vspace*{1cm} 
\centerline{\epsfbox{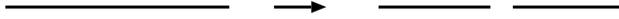}}
\caption{\small splitting of an open string}
\label{open2}
\end{figure}

\begin{figure}
\vspace*{1cm} 
\centerline{\epsfbox{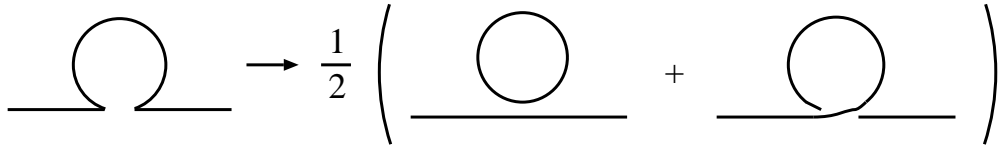}}
\caption{\small splitting and twisting of an open string}
\label{open1}
\end{figure}

\begin{figure}
\vspace*{1cm} 
\centerline{\epsfbox{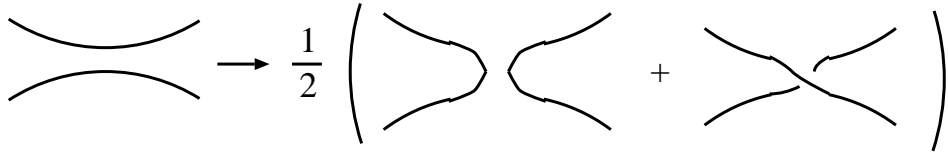}}
\caption{\small joining of two open strings: case one}
\label{openopen1}
\end{figure}

\beqn
\bullet
(\ref{loopeq2}) &\Rightarrow&  \;\;0 =
(1)\; \mbox{\underline{kinetic term (Fig. \ref{openkin1},\ref{open2})}}
 + (2)\; \mbox{\underline{splitting and twisting (Fig.\ref{open1} )}} \\
&+& (3)\;\mbox{\underline{joining with a closed string
 (Fig. \ref{openclosed})}}
+(4)\;\mbox{\underline{joining with an open string (Fig. \ref{openopen1})}}
\;\;.
\nonumber 
\eeqn
  Here
\beqn
 &(1)& = 
\frac{1}{g^2} \left\langle \left( \delta_{X} \Psi[(1); X^r]  \right) 
\Phi[(1)] \cdots \Phi[(N)]  \Psi[(2)] \cdots \Psi[(L)] \right\rangle \;\;, \\
 &(2)& = \left\langle \right.
 \left( -\frac{i}{2}
\sum_{l=l^{(1)}_{0}}^{l^{(1)}_{1}} K^{(1)}_{l} \right) \nonumber\\
&&\left\{
\left( \Lambda^{(1)'}  \Pivector_{(f^{(1)'})} \right) F
U[k^{(1)}_{.},\zeta^{(1)}_{.};l^{(1)}_{2},l^{(1)}_{1}+1]
U[k^{(1)}_{.},\zeta^{(1)}_{.};l,l^{(1)}_{0}]
  \left(\Lambda^{(1)} \Pivector_{(f^{(1)})} \right)
\Phi[k^{(1)}_{.},\zeta^{(1)}_{.};l^{(1)}_{1},l+1]\right.\nonumber\\
 &\pm&  \left. \!\!\!\!\!\! \left( \Lambda^{(1)'} \Pivector_{(f^{(1)'})} \right) F
U[k^{(1)}_{.},\zeta^{(1)}_{.};l^{(1)}_{2},l^{(1)}_{1}+1]
U[\mp k^{(1)}_{.},\mp \zeta^{(1)}_{.};l+1,l^{(1)}_{1}]
U[k^{(1)}_{.},\zeta^{(1)}_{.};l,l^{(1)}_{0}]
  \left( \Lambda^{(1)} \Pivector_{(f^{(1)})} \right) \right\}\nonumber\\
&&\hspace*{4cm} \left. 
\Phi[(1)] \cdots \Phi[(N)]  \Psi[(2)] \cdots \Psi[(L)] \right\rangle
\nonumber\\
&+&  \left\langle \right.
\left( -\frac{i}{2}
\sum_{l=l^{(1)}_{1}+1}^{l^{(1)}_{2}} K^{(1)m}_{l} \right) \nonumber\\
&& \left\{
  \left( \Lambda^{(1)'}  \Pivector_{(f^{(1)'})}  \right) F
U[k^{(1)}_{.},\zeta^{(1)}_{.};l^{(1)}_{2},l+1]
U[k^{(1)}_{.},\zeta^{(1)}_{.};l^{(1)}_{1},l^{(1)}_{0}]
  \left( \Lambda^{(1)} \Pivector_{(f^{(1)})} \right) 
\Phi[k^{(1)}_{.},\zeta^{(1)}_{.};l,l^{(1)}_{1}+1]\right.\nonumber\\
&+& \left.\!\!\!\!\!\! \left( \Lambda^{(1)'} \Pivector_{(f^{(1)'})} \right)   FU[k^{(1)}_{.},\zeta^{(1)}_{.};l^{(1)}_{2},l+1]
U[\mp k^{(1)}_{.},\mp \zeta^{(1)}_{.};l^{(1)}_{1}+1,l]
U[k^{(1)}_{.},\zeta^{(1)}_{.};l^{(1)}_{1},l^{(1)}_{0}]
  \left( \Lambda^{(1)} \Pivector_{(f^{(1)})} \right) \right\} \nonumber\\
&&\hspace*{4cm} \left. 
\Phi[(1)] \cdots \Phi[(N)]  \Psi[(2)] \cdots \Psi[(L)] \right\rangle \;\;, \\
 &(3)& =
\left\langle \right.
 \left( -\frac{i}{2}\sum_{b=1}^{N}
\sum_{n=n^{(b)}_{0}}^{n^{(b)}_{1}} P^{(b)}_{n} \right)  \nonumber \\
&& \left\{
 \left( \Lambda^{(1)'}  \Pivector_{(f^{(1)'})} \right) F
U[k^{(1)}_{.},\zeta^{(1)}_{.};l^{(1)}_{2},l^{(1)}_{1}+1]
U[p^{(b)}_{.},\eta^{(b)}_{.};n,n+1]
U[k^{(1)}_{.},\zeta^{(1)}_{.};l^{(1)}_{1},l^{(1)}_{0}]
 \left( \Lambda^{(1)}  \Pivector_{(f^{(1)})} \right) 
\right.  \nonumber\\
&\mp& \left.\!\!\!\!\!\!\left( \Lambda^{(1)'} \Pivector_{(f^{(1)'})} \right) F
U[k^{(1)}_{.},\zeta^{(1)}_{.};l^{(1)}_{2},l^{(1)}_{1}+1]
U[\mp p^{(b)}_{.},\mp \eta^{(b)}_{.};n+1,n]
U[k^{(1)}_{.},\zeta^{(1)}_{.};l^{(1)}_{1},l^{(1)}_{0}]
\left(  \Lambda^{(1)} \Pivector_{(f^{(1)})} \right) \right\} \nonumber\\
&&\hspace*{4cm} \left. 
\Phi[(1)] \cdots \hat{b} \cdots \Phi[(N)]  
\Psi[(2)] \cdots \Psi[(L)] \right\rangle \;\;, \\
  &(4)& =   \left\langle \right.
\left( -\frac{i}{2}\sum_{b=2}^{L} \sum_{l=l^{(b)}_{0}}^{l^{(b)}_{1}}
 K^{(b)}_{l}
  \right) \nonumber\\ 
&&  \left\{  \left( \Lambda^{(1)'}  \Pivector_{(f^{(1)'})} \right) F
U[k^{(1)}_{.},\zeta^{(1)}_{.};l^{(1)}_{2},l^{(1)}_{1}+1]
U[k^{(b)}_{.},\zeta^{(b)}_{.};l,l^{(b)}_{0}]
 \left( \Lambda^{(b)} \Pivector_{(f^{(b)})} \right) 
\right.  \nonumber\\
&& \left( \Lambda^{(b)'} \Pivector_{(f^{(b)'})} \right)
U[k^{(b)}_{.},\zeta^{(b)}_{.};l^{(b)}_{1},l+1]
U[k^{(1)}_{.},\zeta^{(1)}_{.};l^{(1)}_{1},l^{(1)}_{0}]
 \left(  \Lambda^{(1)}  \Pivector_{(f^{(1)})} \right) \nonumber\\
&\pm&   \left( \Lambda^{(1)'}  \Pivector_{(f^{(1)'})} \right) F
U[k^{(1)}_{.},\zeta^{(1)}_{.};l^{(1)}_{2},l^{(1)}_{1}+1]
U[\mp k^{(b)}_{.},\mp \zeta^{(b)}_{.};l+1,l^{(b)}_{1}]
 \left( \Lambda^{(b)'}  \Pivector_{(f^{(b)'})} \right) \nonumber\\
&\times& \left.  \left( \Lambda^{(b)} \Pivector_{(f^{(b)})} \right) F
U[\mp k^{(b)}_{.},\mp \zeta^{(b)}_{.};l^{(b)}_{0},l]
U[k^{(1)}_{.},\zeta^{(1)}_{.};l^{(1)}_{1},l^{(1)}_{0}]
 \left( \Lambda^{(1)}  \Pivector_{(f^{(1)})} \right)  \right\} \nonumber\\
&&\hspace*{3cm} \left. 
\Phi[(1)] \cdots \Phi[(N)]  
\Psi[(2)] \cdots \hat{b} \cdots \Psi[(L)] \right\rangle \;\;.
\label{loopeq2complete}
\eeqn
  The form of  $\delta_{X} \Psi[(1); X^r]$ is similar in spirit
 to that of  $\delta_{X} \Phi[(1); X^r]$. Space permits us to  write  this
 explicitly only for the case $X^r =v_{m}^{r}$ in Appendix B.


\begin{figure}
\vspace*{1cm} 
\centerline{\epsfbox{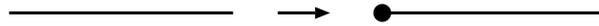}}
\caption{\small infinitesimal deformation of an open string: case two}
\label{openkin2}
\end{figure}

\begin{figure}
\vspace*{1cm} 
\centerline{\epsfbox{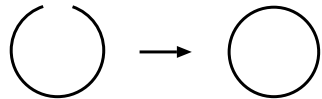}}
\caption{\small open-closed transition}
\label{open-closed}
\end{figure}

\begin{figure}
\vspace*{1cm} 
\centerline{\epsfbox{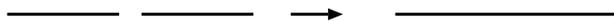}}
\caption{\small joining of two open strings: case two}
\label{openopen2}
\end{figure}

\beqn
\bullet
(\ref{loopeq3}) &\Rightarrow&  \;\;0  =
(1)\; \mbox{\underline{kinetic term (Fig. \ref{openkin2})}} 
 + (2)\; \mbox{\underline{open-closed transition 
(Fig. \ref{open-closed})}} \\
&+& (3)\;
 \mbox{\underline{joining with an open string (Fig. \ref{openopen2})}} \;\;.
\nonumber
\eeqn
Here
\beqn
 &(1)& = 
\frac{1}{g^2} \left\langle \left( \delta_{\Zvector} \Psi[(1);\Zvector] 
 \right) 
\Phi[(1)] \cdots \Phi[(N)]  \Psi[(2)] \cdots \Psi[(L)] \right\rangle \;\;, \\
 &(2)&  =
\delta_{f f^{(1)}} \Lambda^{(1)} \left\langle 
\Phi[k^{(1)}_{.},\zeta^{(1)}_{.};l^{(1)}_{1},l^{(1)}_{0}]
\Phi[(1)] \cdots \Phi[(N)]  \Psi[(2)] \cdots \Psi[(L)] \right\rangle \;\;,\\
 &(3)& =
 \left\langle  \right. \sum_{b=2}^{L}
 \left\{ \delta_{f f^{(b)'}} \Lambda^{(b)'} \right.   \nonumber \\
 &\times& \left(   \Lambda^{(1)} \Pivector_{(f^{(1)})} \right)
FU[\mp k^{(1)}_{.},\mp \zeta^{(1)}_{.};l^{(1)}_{0},l^{(1)}_{1}]
U[k^{(b)}_{.},\zeta^{(b)}_{.};l^{(b)}_{1},l^{(b)}_{0}]
\left(  \Lambda^{(b)} \Pivector_{(f^{(b)})} \right)  \nonumber\\
   &\pm& \left.   \delta_{f f^{(b)}} \Lambda^{(b)}
 \left(  \Lambda^{(b)'} \Pivector_{(f^{(b)'})} \right)
FU[k^{(b)}_{.},\zeta^{(b)}_{.};l^{(b)}_{1},l^{(b)}_{0}]
U[k^{(1)}_{.},\zeta^{(1)}_{.};l^{(1)}_{1},l^{(1)}_{0}]
  \left( \Lambda^{(1)} \Pivector_{(f^{(1)})} \right)  \right\} \nonumber\\
&&\hspace*{3cm} \left. 
\Phi[(1)] \cdots \Phi[(N)]  
\Psi[(2)] \cdots \hat{b} \cdots \Psi[(L)] \right\rangle\;\;.
\label{loopeq3complete}
\eeqn
  The form of  $\left( \delta_{\Zvector} \Psi[(1);\Zvector] \right)$
 is in Appendix B.


\begin{figure}
\vspace*{1cm} 
\centerline{\epsfbox{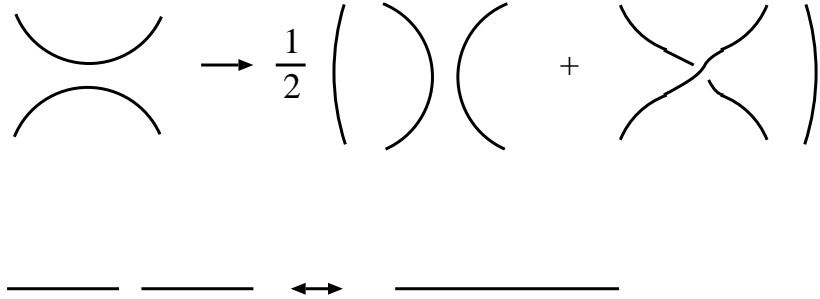}}
\caption{\small two kinds of elementary local processes}
\label{elementaryprocess}
\end{figure}

  We have checked that all of the terms in (\ref{loopeq1complete}) $\sim$
 (\ref{loopeq3complete})  are expressed by the closed and open loops
 $\Phi$ and $\Psi$ and their derivatives with respect to the sources
 introduced.
For example, the expression
$\psivector^{\ast} \cdot  \bar{\sigma}^m
U[p^{(1)}_{.},\eta^{(1)}_{.};n^{(1)}_{1},n^{(1)}_{1}+1] \psivector$
in eq. (\ref{eq:app30}) for $\left( \delta_{X} \Phi[(1); X^r] \right)$
in  eq. (\ref{eq:loop1kin}) is represented as
\beq 
\sum_{f=1}^{2n_f}\bar{\sigma}^{\dot{\alpha}\alpha m} 
\frac{\partial}{\partial\bar{\theta}^{'\dot{\alpha}}}
\frac{\partial}{\partial\theta^{\alpha}} \Psi_{f f}[p^{(1)}_{.},
\eta^{(1)};n^{(1)}_{1},n^{(1)}_{1}+1;\Lambda',\Lambda]\;\;.
\eeq
In this sense, the set of loop equations we have derived is closed. 
 It is noteworthy that all of the terms in the above loop equations
are either an infinitesimal deformation of a loop or a consequence from 
  the two elementary local processes of loops which are illustrated 
   in Fig. \ref{elementaryprocess}.

 It is interesting to discuss the system of loop equations we have derived 
in the light of string field theory.  In addition to the lightcone
superstring field theory constructed earlier in \cite{GSSFT}, there is 
now gauge invariant string field theory for closed-open bosonic system
\cite{KT}. We find that the types of the interaction terms of our equations
are in complete agreement with the interaction vertices seen in 
\cite{GSSFT}  and
the second paper of \cite{KT}. In particular, Figures $2\sim5, 7\sim9,
11\sim12$  for the interactions of our equations are in accordance with
  $U, V_{\infty}, V_{3}^{c}, U_{\Omega}, V_{3}^{0}, V_{\alpha}, V_{4}^{0}$
  of \cite{KT}. 
   While BRS invariance  determines  the coefficients of the interaction
 vertices in  \cite{KT}, the (bare) coefficients are already determined
 in our case  from the first quantized action.  This may give us insight
 into  properties of the model wchich are not revealed.

\section{Linearized loop equations and  a free string}
  Let us consider the all three loop equations
eqs. (\ref{loopeq1}), ({\ref{loopeq2})  and (\ref{loopeq3}) in the linearized
 approximation,  namely, ignoring  the joining and splitting of the loops. 
  Let us first introduce a variable conjugate to $p_{A n}$ or $k_{A n}$
  and that to $\eta_{n}$ or $\zeta_{n}$  by
\beqn
\hat{X}^{A}_{n} &=& i \frac{\delta}{\delta p_{A n}} \; \mbox{or} \;
                 \; i \frac{\delta}{\delta k_{A n}}  \;\;, \\
\hat{\Psi}_{n} &=&  i \frac{\delta}{\delta \eta_{n}} \; \mbox{or} \;
               \;  i \frac{\delta}{\delta \zeta_{n}}\;\;.
\eeqn
By acting $\hat{X}^{A}_{n}$ and $\hat{\Psi}_{n}$ on a loop, we obtain
   respectively  an  operator insertion
of $v^{A}$ and that of $\Psi$ at point $n$  on the loop.

Now consider   eq. (\ref{loopeq1}) and  eq. (\ref{loopeq2}) 
for  the case $X^r=v_{M}^{r}$,   multiplying them by $p_{nM}^{(1)}$ and
$k_{nM}^{(1)}$ respectively.     Consistency requires that,
 for these terms, we must take into account
 the term from the interactions which represents splitting of 
  a loop  with infinitesimal length.
  This in fact occurs when
 the splitting point $n$ coincides with the point $n_{1}^{(1)}$ 
at which $T^r$ is inserted.
We obtain
\beqn
0&=&\frac{1}{g^2} p_{nM}^{(1)}
\left\langle  \left( \delta_{X} \Phi[(1); v_{M}^{r}]  \right) 
\Phi[(2)] \cdots \Phi[(N)]  \Psi[(1)] \cdots \Psi[(L)]
 \right\rangle \nonumber\\
&&-\frac{i}{2}2k p_{n}^{(1)2}
\left\langle  \Phi[(1)] 
\Phi[(2)] \cdots \Phi[(N)]  \Psi[(1)] \cdots \Psi[(L)] \right\rangle  \;\;,\\
0&=&\frac{1}{g^2} k_{nM}^{(1)}
\left\langle  \left( \delta_{X} \Psi[(1); v_{M}^{r}]  \right) 
\Phi[(1)] \cdots \Phi[(N)]  \Psi[(2)] \cdots \Psi[(L)] \right\rangle \nonumber\\
&&-\frac{i}{2}2k k_{n}^{(1)2}
\left\langle  \Phi[(1)] 
\Phi[(2)] \cdots \Phi[(N)]  \Psi[(1)] \cdots \Psi[(L)] \right\rangle \;\;.
\eeqn
These equations lead to the half of the Virasoro conditions \cite{FKKT}:
\beqn
0&=&(p_{n}^{(1)2}+\hat{X}_{n}^{(1)'2}+\mbox{(fermionic part)}) \nonumber\\
&&\left\langle  \Phi[(1)] 
\Phi[(2)] \cdots \Phi[(N)]  \Psi[(1)] \cdots \Psi[(L)] \right\rangle  \;\;, \\
0&=&(k_{n}^{(1)2}+\hat{X}_{n}^{(1)'2}+\mbox{(fermionic part)}) \nonumber\\
&&\left\langle  \Phi[(1)] 
\Phi[(2)] \cdots \Phi[(N)]  \Psi[(1)] \cdots \Psi[(L)] \right\rangle \;\;, 
\eeqn
where $\prime$  implies  taking  a difference between  two adjacent
 points $n$  and $n+1$.
The reparametrization invariance of the Wilson loops leads to
the remaining half of the Virasoro conditions:
\beqn
0&=&(p_{n}^{(1)M}\hat{X}_{nM}^{(1)'}+\mbox{(fermionic part)}) \nonumber\\
&&\left\langle  \Phi[(1)] 
\Phi[(2)] \cdots \Phi[(N)]  \Psi[(1)] \cdots \Psi[(L)] \right\rangle \;\;, \\
0&=&(k_{n}^{(1)M}\hat{X}_{nM}^{(1)'}+\mbox{(fermionic part)}) \nonumber\\
&&\left\langle  \Phi[(1)] 
\Phi[(2)] \cdots \Phi[(N)]  \Psi[(1)] \cdots \Psi[(L)] \right\rangle \;\;.
\eeqn

Next, let us consider  eq. (\ref{loopeq3}), ignoring joining and splitting
 of the loops.  Again consistency appears to require that
 we drop the cubic terms consisting of $\Qvector$ and $\Qvector^{\ast}$
   in   $\delta_{\Zvector} \Psi[(1);\Zvector]$ .
  To write explicitly, the following expression   must vanish
\beqn
&& \left\{ \Qvector^{\ast}_{(f)}(v_{\nu}v^{\nu}+[\Phi_I,\Phi^I])
+(\Qvector \Sigma)_{(f)}F[\Phi_{2}^{\dagger},\Phi_{3}^{\dagger}]
+(\Qvector^{\ast}M^2)_{(f)}+2 (\Qvector^{\ast}M)_{(f)}v_4
- i \sqrt{2} \psivector^{\ast}_{(f)}\bar{\lambda}   \right.  \nonumber\\
&& \left. -\sqrt{2}(\psivector \Sigma)_{(f)} F \psi_{\Phi_1} \right\} 
U[k^{(1)}_{.},\zeta^{(1)}_{.};l^{(1)}_{1},l^{(1)}_{0}] 
(\Lambda^{(1)} \Pivector_{f^{(1)}}) \approx  0 
  \label{linearizedloopeq1} \;\;\;, \\
&& \left\{ \psivector^{\ast}_{(f)} \bar{\sigma}^{m} v_{m}
+i \sqrt{2} \Qvector_{(f)}^{\ast} \lambda 
+ \left( \psivector  \Sigma F (\sqrt{2}\Phi_1 + M) \right)_{(f)}  \right.
   \nonumber \\
&&\left. +\sqrt{2} (\Qvector  \Sigma F \psi_{\Phi_1})_{(f)} \right\}
U[k^{(1)}_{.},\zeta^{(1)}_{.};l^{(1)}_{1},l^{(1)}_{0}] 
(\Lambda^{(1)} \Pivector_{f^{(1)}}) \approx  0   \;\;\;,
\label{linearizedloopeq2}
\eeqn
when inserted in 
\beq
\left\langle  \Phi[(1)] 
\Phi[(2)] \cdots \Phi[(N)] \Psi[(2)] \cdots \Psi[(L)]  \right\rangle \;\;.
\eeq

  As we stated before, the lefthand sides of eqs.
(\ref{linearizedloopeq1}) and (\ref{linearizedloopeq2}) are expressible
as an open loop with some functions of $\hat{X}^{A}_{l^{(1)}_{1}}$  and
$\hat{\Psi}_{l^{(1)}_{1}}$ acting on the loop. Let us see by inspection how
eqs. (\ref{linearizedloopeq1}) and (\ref{linearizedloopeq2}) are satisfied by
 the source functions alone.
Consider  the following configuration of $\hat{X}_{n}$ and $\hat{\Psi}_{n}\;,$
$n= l^{(1)}_{1}$:
\beqn
  && \hat{X}^{\mu}    \approx 0  \;\; \mbox{for}\;\; \mu = 0,1,2,3,7 \;\;,
  \;\;\;\hat{X}^{4}=\pm m_f   \;\;\;.  \label{boundary1}   \\
&& \hat{X}^{'I} \approx 0  \;\; \mbox{for}\;\;  I = 5,6,8,9 
 \label{boundary2}  \\
  &&   \hat{\Gamma}_{3} \hat{\Psi}  \approx  -\hat{\Psi} \;\;\;,
\label{boundary3}
\eeqn
where $\hat{\Gamma}_{3}  \equiv \Gamma_5 \Gamma_6 \Gamma_8 \Gamma_9$.  
Again  these equations  should be understood in the sense of 
   an insertion at the end point of the open loop.

  Eqs. (\ref{boundary1}), (\ref{boundary2}), and (\ref{boundary3}) tell us
 that the open loop $\Psi[(1)]$   obeys the Dirichlet boundary
conditions for $0,1,2,3,4,7$ directions and the Neumann boundary 
conditions for $5,6,8,9$ directions. Eq. (\ref{boundary3}) is seen
 in \cite{EMM}.
  Note that eq. (\ref{boundary3}) is equivalent to
\beq
\hat{\lambda} \approx   \hat{\bar{\lambda}} \approx  \hat{\psi}_{\Phi_1}
 \approx \hat{\bar{\psi}}_{\Phi_1}  \approx  0 \;\;.
\eeq 
  Also note that
\beq
[\Phi_i,\Phi_j] \approx   \hat{X'}_i \hat{X}_j \; \mbox{or} \;
 \hat{X'}_j \hat{X}_i \;\;.
\eeq

 We find that  the configuration given by eqs. (\ref{boundary1}),
(\ref{boundary2}) and (\ref{boundary3}) solves the linearized loop equations
(\ref{linearizedloopeq1}), (\ref{linearizedloopeq2}).
This configuration clearly tells us the existence of $n_f$ $D3$ branes
and their mirrors  each of which is  at a distance $\pm m_f$ away from the
 orientifold surface in the fourth direction.
 There has been positive evidence in favour of this both from
 the connection of the T dualized ( the $4D$ worldvolume  gauge theory)
 representation of our model \cite{IT1,IT2} with Sen's scaling limit
 \cite{Sen} for F theory \cite{V}  and from the
 configuration emerging from the fermionic integration \cite{IM,CIK}.
  ( See also \cite{fermi}.)
  The result in this section consolidates our picture.

\section{Discussion}
  We have been able to formulate the $USp$ matrix model in the second
quantized form in which the manybody effects of the model as string theory
 are manifest.   From our discussion, it is
clear that the closed and open Wilson loop variables  serve as string fields.
It is satisfying to see that the linearized equations translate into the
classical Virasoro condition of
  the closed loops/string fields and the boundary conditions of the open
  loops/string fields.  It is encouraging to us for a further pursuit of
  the model that the simple completeness relation of the $usp$  Lie algebra 
 is able to capture the complete set of the joining and splitting 
 interactions required.

  While our paper supplies several satisfactory features of the model
   as unified theory of all forces including gravity  and matter,
  it provides us  with  a host of open questions many of which are shared by
  the type $IIB$ case.   Let us discuss some of them.  The theory  is  still
 formulated in bare variables and the proper scaling limit is yet to be
 determined.  This limit
in the $USp$ case is closely related to the problem of the field/loop
redefinition and therefore relative strengths of the string interactions
among the closed and open string fields
and that of the $type I$-heterotic duality\cite{PW} of the $USp$ matrix model. 
  A related but different question is
how, given a model, we find  perturbative vacuum on which string
 perturbation theory is based.  This is a nontrivial problem in reduced
 models as perturbative vacuum  is neither the true vacuum realized by the
 scaling limit  nor simple theory of loops as bare variables which
  ignores  the joining and splitting of the loops.
   As we discussed at the beginning, the connection of the reduced model
 action in the large $k$ limit with the first quantized string action in
  the Schild gauge  ensures that string perturbation theory is somewhere
 in the model. The nonrenormalizability of the worldsheet action and the
 absence of free field technique, however, prevent us from the direct study.
 
 Turning to physical consequences,
 the reduced matrix model, in particular, their second quantized formulation
  provides an opportunity to answer questions which are difficult to
 address in the conventional first quantized string theory.
   These are, for example,
 the size and the shape of spacetime which the model predicts and the issue
 of spontaneous breaking of gauge symmetry. These can be studied
  within the model by numerical  as well as analytical method.

\section{Acknowledgements}
      We thank  Satoshi Iso for a helpful discussion on this subject.

\newpage

\appendix

\section{}

\subsection{the action of the $USp$ matrix model}
  The action of  the  $USp(2k)$ reduced matrix model can be
 obtained from the dimensional reduction of
 ${\cal N}=2, d=4$ $USp(2k)$ supersymmetric gauge theory with one
hypermultiplet in the antisymmetric representation and $n_{f}$ hypermultiplets
in the fundamental representation. This makes manifest the presence of the
eight dynamical supercharges.
 In the ${\cal N}=1$ superfield notation  with
  spacetime dependence all dropped, we have a vector superfield $V$ and a
  chiral superfield $\Phi \equiv \Phi_{1}$ which are $usp$ Lie algebra valued
\beqn
\label{eq:adj}
  V^{t} = - FV F^{-1} \;\;, \;\; \Phi^{t} = - F \Phi F^{-1}  \;\;,
 \;\; V^{\dagger} = V \;\;, \;\;  {\rm with} \;\;  F =
    \left(
 \begin{array}{cc}
   0 & I  \\
   -I & 0  
  \end{array}   \right)\;\;,
\eeqn
 and the two chiral superfields $\Phi_{I} \;,\; I =2,3$ in the antisymmetric
 representation which obey
\beqn
\label{eq:asym}
  \Phi_{I}^{t}=  F \Phi_{I} F^{-1} \;\;\; {\rm for\; I=2,3}\;\;.
\eeqn 
  We will suppress the $USp$ indices  in the rest of our discussion.

  It is often expedient to introduce the projector acting on $U(2k)$ matrices:
\beqn
\label{eq:projector}
 \hat{\rho}_{\mp} \bullet  \equiv \frac{1}{2} \left( \bullet \mp F^{-1}
 \bullet^{t}  F \right) \;\;\;.
\eeqn
 The action  of  $\hat{\rho}_{-}$   and that of  $\hat{\rho}_{+}$   take
 any $U(2k)$ matrix  into  the matrix lying in the adjoint representation
 of $USp(2k)$  and that in the antisymmetric representation respectively.
   We can  therefore write
$V =  \hat{\rho}_{-} \ud{V},\;\;   \Phi_{1} = \hat{\rho}_{-}
 \ud{\Phi_{1}} ,\;\;
   \Phi_{I}  =   \hat{\rho}_{+} \ud{\Phi_{I}} , \;\;  I = 2,3 ,$
 where the  symbols with underlines  lie in the adjoint representation
 of $U(2k)$.
  The total action is written as
\beqn
\label{eq:vvaym}
 S &=&
 \frac{1}{4 g^{2}} \;  Tr \left(
 \int d^2 \th W^{\a} W_{\a} + h.c. +
4 \int d^2 \th d^2 \thb \Phi^{\dg}_{I} e^{2V} \Phi_{I} e^{-2V} \right) \\
  &+&  \frac{1}{g^{2}} \sum_{f=1}^{n_f}
  \int d^2 \th d^2 \thb
\left( Q_{(f)}^{*} \left( e^{2V} \right) Q_{(f)}
+  \tilde{Q}_{(f)} \left( e^{-2V} \right) \tilde{Q}_{(f)}^{*} \right)
  + \frac{1}{g^{2}} \left( \int d^2\th W ( \th) + h.c. \;\right)\;\;\;,
\nonumber
\eeqn
 where  the superpotential
\beqn
 W ( \th)  = \sqrt{2} Tr \left(
   \Phi_{1} \left[ \Phi_{2},  \Phi_{3} \right] \right) 
        +  \sum_{f=1}^{n_{f}} \left( m_{(f)} \tilde{Q}_{(f)} Q_{(f)}
        + \sqrt{2} \tilde{Q}_{(f)} \Phi_{1}Q_{(f)} \right) \;\;\;. 
\eeqn
 
  To render the action to its component form, let us first list some formulas.
$W_{\alpha} = -\frac{1}{8} \Db \Db e^{-2V} D_{\a} e^{2V},\;
\Phi_{I} = \Phi_{I} + \sqrt{2} \th \psi_{\Phi_{I}} + \th \th F_{\Phi_{I}}
,$  $\; Q= Q + \sqrt{2} \th \psi_{Q} + \th \th F_{Q}$,
$\;V = - \th \s^m \thb v_m + i \th \th \thb \lb - i \thb \thb \th
\l + \half \th \th \thb \thb D $,
$\;D_{\a} = \frac{\p}{\p \th^{\a}},$ $\;\bar{D}_{\dot{\a}} =
  -\frac{\p}{\p \thb^{\dot{\a}}} $.
 Solving the equation for the $D$ term, we obtain
\beqn
  D = \left[ \Phi^{\dagger}_{I}, \Phi_{I}\right] -  \hat{\rho}_{-}
 \sum_{f=1}^{n_{f}} \left( Q_{(f)}  Q_{(f)}^{*} -
 \tilde{Q}_{(f)}^{*}   \tilde{Q}_{(f)} \right)  \;\;,
\eeqn
where we have placed the $USp$ vectors $Q_{(f)},\; \tilde{Q}_{(f)}$ and their
  complex conjugates in the form of dyad.  The $F$ terms  are such that
\beqn 
 - \delta W = \sum_{I=1,2,3} tr F^{\dagger}_{\Phi_{I}} \delta \Phi_{I}
      + F^{*}_{Q_{(f)}} \delta  Q_{(f)} 
+ F^{*}_ {\tilde{Q}_{(f)}}  \delta \tilde{Q}_{(f)} \;\;\;.
\eeqn
  Explicitly
\beqn
\label{eq:F}
  F^{\dagger}_{\Phi_{1}} &=&  - \sqrt{2} \left[ \Phi_{2}, \Phi_{3}\right]
  - \sqrt{2} \hat{\rho}_{-} \left( \sum_{f=1}^{n_{f}} Q_{(f)}
 \tilde{Q}_{(f)} \right),\;
 F^{\dagger}_{\Phi_{2}} =  - \sqrt{2} \left[ \Phi_{3}, \Phi_{1}\right],
 \;\; 
 F^{\dagger}_{\Phi_{3}} = - \sqrt{2} \left[ \Phi_{1}, \Phi_{2}\right] \;,
  \nonumber \\
 F^{*}_{Q_{(f)}} &=& - \left( m_{(f)} \tilde{Q}_{(f)} + \sqrt{2}
 \tilde{Q}_{(f)} \Phi_{1} \right) \;,\;\;
  F^{*}_{\tilde{Q}_{(f)}} = - \left( m_{(f)} Q_{(f)} + \sqrt{2} \Phi_{1} 
 Q_{(f)} \right) \;\;.
\eeqn
 As for the Yukawa couplings, they can be read off from
\beqn
 \delta^{2} W \equiv  \sum_{A,B} \frac{\partial^{2}W}{\partial A \partial B}
\delta A \delta B \;\;,
\eeqn
  where the summmation indices $A,\;B$ are over all chiral superfields
 $\Phi_{I}\; I=1,2,3,$ and $Q_{(f)}, \tilde{Q}_{(f)}\;, \;
 f= 1, \cdots n_{f}$.
  The component expression for the total action is
\beqn
 S &=& 
\frac{1}{g^2} Tr \left\{- \frac{1}{4} v_{m n} v^{m n} -
[\cD_{m}, \Phi_{I}]^{\dagger} [ \cD^{m}, \Phi_{I} ]
- i \lambda \s^{m} [ \cD_{m} , \overline{\lambda} ]  
- i \overline{\psi}_{\Phi_{I}} \overline{\s}^{m} [\cD_{m} , \psi_{\Phi_{I} } ]
- i \sqrt{2} [ \l , \psi_{\Phi_{I}} ] \Phi^{\dagger}_{I} \right.  \nonumber \\
 & &\left.  - i \sqrt{2}
[ \overline{\lambda}   , \overline{\psi}_{\Phi_{I} } ] \Phi_{I}
 \right\} \nonumber \\
 &+& \frac{1}{g^2} \sum_{f=1}^{n_f}
 \left\{
- ( \cD_{m} Q_{(f)} )^{*} (\cD^{m} Q_{(f)} )
- i {\overline{\psi}}_{Q (f)}  \overline{\s}^{m}
\cD_{m} \psi_{Q (f)}
+ i \sqrt{2} Q_{(f)}^{*} \l \psi_{Q (f)}
- i \sqrt{2} {\overline{\psi}}_{Q (f)}  \overline{\lambda} Q_{(f)}
 \right\}  \nonumber \\
&+& \frac{1}{g^2} \sum_{f=1}^ {n_f} \left\{
- ( \cD_{m} {\tilde{Q}}_{(f)} ) (\cD^{m} {\tilde{Q}}_{(f)} )^{*}-
 i {\overline{\psi}}_{{\tilde{Q}} (f)}  \overline{\s}^{m}
\cD_{m}^{*} \psi_{{\tilde{Q}} (f)}
- i \sqrt{2} \psi_{{\tilde{Q}} (f)} \l  \tilde{Q}_{(f)}^{*}  
+ i \sqrt{2}   {\tilde{Q}}_{(f)}  \overline{\lambda}
 {\overline{\psi}}_{{\tilde{Q}} (f)}  \right\}   \nonumber \\
 &-& \frac{1}{g^2} Tr \left(
\frac{1}{2} D D   + F_{\Phi_{I}}^{\dagger} F_{\Phi_{I}} \right)
 -\frac{1}{g^{2}} \left( \sum_{f=1}^{n_f} F_{Q_{(f)}} F_{Q_{(f)}}^{*}
 + \sum_{f=1}^{n_f} F_{\tilde{Q}_{(f)}} F_{\tilde{Q}_{(f)}}^{*}
\right)    \nonumber \\
&-&\frac{1}{g^2} \left( \sum_{A,B} \frac{\partial^{2}W}{\partial A \partial B}
 \psi_{A} \psi_{B} + h.c.\right) \;\;.
\eeqn
   Here  $\cD_{m} = i v_{m}$ in the fundamental representation.

 Let us denote by  $S_{0}$
 the part in $S$ which does not contain the fundamental
 hypermultiplet. 
  We split the total action into
\beqn
  S= S_{0} + \Delta S \;\;\;.  \nonumber 
\eeqn
  The part  $S_{0}$ is expressible in terms of the type $II B$ matrix
 model.  This is stated as
\beqn
S_{0} ( v_{m} , \Phi_{I},
 \lambda, \psi_{\Phi_{I}},
\bar{\Phi}_{I}, \bar{\lambda},  \bar{\psi}_{\Phi_{I} }, )
 = S_{IIB}( \hat{\rho}_{b\mp}\ud{v}_{M}, \hat{\rho}_{f\mp}\ud{\Psi}  ) \;\;.
\label{eq:equiv}
\eeqn
 Here
\beqn
   S_{IIB}(\ud{v}_{M},  \ud{\Psi} )  =
  \frac{1}{g^{2}} Tr \left( \frac{1}{4} \left[\ud{v}_{M},
 \ud{v}_{N} \right]
\left[ \ud{v}^{M}, \ud{v}^{N} \right] - \frac{1}{2}
\bar{\ud{\Psi}} \Gamma^{M} \left[ \ud{v}_{M},
\ud{\Psi} \right] \right) \;\;,
\eeqn
and
\beqn
\label{eq:phiv}
\Phi_{i} &=& \frac{1}{\sqrt{2}} \left( v_{3+i} +i v_{{6+i}}
 \right) \;\;, \;  \nonumber \\   {\rm and} \;\;\;\;\;\;\;
\Psi  &=&  \left(
 \lambda , 0, \psi_{\Phi_{1}}, 0 ,\psi_{\Phi_{2}}, 0  ,  \psi_{\Phi_{3}}, 0 ,
  0, \bar{\lambda}, 0,
 \bar{\psi}_{\Phi_{1}}, 0,
 \bar{\psi}_{\Phi_{2}},  0, \bar{\psi}_{\Phi_{3}} \right)^t \;\;.
\eeqn
 This latter one $\Psi$ is a thirty two component Majorana-Weyl spinor
satisfying
\beqn
  C \bar{\Psi}^t = \Psi  \;
 , \; \Gamma_{11} \Psi =  \Psi \;\;\;.
\label{eq:majorana-weyl con}
\eeqn
With regard to eqs. (\ref{eq:phiv}) and (\ref{eq:majorana-weyl con}),
the same is true for objects with underlines.
 The ten dimensional gamma matrices have been denoted by 
 $\Gamma^{M}$.
  The projector  $\hat{\rho}_{b\mp} $ is a diagonal matrix with
 respect to Lorentz indices
  while $\hat{\rho}_{f\mp} $ is  to
 spinor indices:
\beqn
 \hat{\rho}_{b\mp}
 &=& diag
(\hat{\rho}_{-},\hat{\rho}_{-},\hat{\rho}_{-},
 \hat{\rho}_{-},\hat{\rho}_{-},\hat{\rho}_{+},
 \hat{\rho}_{+},\hat{\rho}_{-}, \hat{\rho}_{+},\hat{\rho}_{+} )
\nonumber \\
 \hat{\rho}_{f\mp}   &=&
\hat{\rho}_{-} 1_{(4)} \otimes
\left( \begin{array}{cccc}
        1_{(2)}& & & \\
               & 0 & & \\
               &   & 1_{(2)} & \\
               &   &         &0
        \end{array}
\right)
+
\hat{\rho}_{+} 1_{(4)} \otimes
\left( \begin{array}{cccc}
        0& & & \\
               & 1_{(2)} & & \\
               &   & 0 & \\
               &   &         & 1_{(2)}
        \end{array}
\right)
 \;\;.
\label{eq:projectors of b f}
\eeqn
 The proof of the equivalence (\ref{eq:equiv}) is sketched here for
  this appendix to be self-contained.  (See also \cite{BSS}).
 The only nontrivial term in the bosonic part of this equivalence
 is 
\beqn
\label{eq:FD}
   \frac{1}{2} Tr D^{(0)} D^{(0)}   + Tr F_{\Phi_{i}}^{\dagger(0)}
 F_{\Phi_{i}}^{(0)}
 = -\frac{1}{4} Tr \left[ v_{r}, v_{r^{\prime}} \right]
  \left[ v^{r}, v^{r^{\prime}} \right] \;\;,
\eeqn
  where $r, r^{\prime}$ run over $ 4 \sim 9$ and the superscript  $(0)$
 implies omission of the parts containing the
 fundamental scalars in eq. (\ref{eq:F}). The left hand side is written as
\beq
\label{eq:prejacobi}
tr \left\{ \frac{1}{2}\left( [ \Phi_{1}^{\dagger}, \Phi_{1} ]^{2}  +
 [\Phi_{2}^{\dagger}, \Phi_{2} ]^{2} +  [ \Phi_{3}^{\dagger}, \Phi_{3} ]^{2}
 \right) +  \sum_{(K,J)} \left( [ \Phi^{\dagger}_{K}, \Phi_{K} ]
 [\Phi^{\dagger}_{J}, \Phi_{J} ]  -2   [\Phi_{K}, \Phi_{J} ]
[ \Phi^{\dagger}_{K}, \Phi^{\dagger}_{J} ]  \right) \right\},
\eeq 
 where the sum $(K,J)$ runs over the pairs $(1,2), (2,3), (3,1)$.
 Using the Jacobi identity, we convert  the first term of the summand into
  another expression. The summand  becomes
$  -[\Phi_{K}, \Phi^{\dagger}_{J} ] [ \Phi^{\dagger}_{K}, \Phi_{J} ]
 -[\Phi_{K}, \Phi_{J} ] [ \Phi^{\dagger}_{K}, \Phi^{\dagger}_{J} ]$.
 Substituting eq. (\ref{eq:phiv}) into eq. (\ref{eq:prejacobi}), we confirm
 eq. (\ref{eq:FD}).
  As for the fermion bilinear, we check  the equivalence eq. (\ref{eq:equiv})
 by finding  an explicit representation of the ten $\Gamma^{M}$
 matrices  in the bases (\ref{eq:phiv}).  They are
\beqn
\Gamma^m&=&\gamma^m \otimes I_8 \;\; \mbox{for}\; m=0,1,2,3, \nonumber\\
&&\gamma^m=\left( \begin{array}{cc}
                        0 & \sigma^m \\
                   \bar{\sigma}^m & 0 
                  \end{array}  \right) , \nonumber\\
\Gamma^I&=&\gamma_5 \otimes \hat{\Gamma}^I \;\; \mbox{for}\; I=4 \sim 9, \nonumber\\
&&\gamma_5=\left( \begin{array}{cc}
                        I_2 & 0 \\
                          0 & -I_2
                  \end{array}  \right),
\label{eq:gamma}
  \;\;\
\eeqn
where 
\beqn
\hat{\Gamma}^4=\left( \begin{array}{cc}
           \Large{0} & \begin{array}{cccc}
                     0 & -i & 0 & 0 \\ 
                     i & 0  & 0 & 0 \\
                     0 & 0  & 0 & 1 \\
                     0 & 0  & -1 &0  \end{array}\\
\begin{array}{cccc}
       0 & i & 0 & 0 \\            
      -i & 0  & 0 & 0 \\
       0 & 0  & 0 & 1 \\
       0 & 0  & -1 &0  \end{array} & \Large{0} \end{array} \right),   
\hat{\Gamma}^5=\left( \begin{array}{cc}
           \Large{0} & \begin{array}{cccc}
                     0 & 0 & -i & 0 \\ 
                     0 & 0 & 0 & -1 \\
                     i & 0 & 0 & 0 \\
                     0 & 1 & 0 & 0  \end{array}\\
\begin{array}{cccc}
       0 & 0 & i & 0 \\            
       0 & 0  & 0 & -1 \\
       -i & 0  & 0 & 0 \\
       0 & 1  & 0 &0  \end{array} & \Large{0} \end{array} \right), \nonumber\\
\hat{\Gamma}^6=\left( \begin{array}{cc}
           \Large{0} & \begin{array}{cccc}
                     0 & 0 & 0 & -i \\ 
                     0 & 0  & 1 & 0 \\
                     0 & -1  & 0 & 0 \\
                     i & 0  & 0 &0  \end{array}\\
\begin{array}{cccc}
       0 & 0 & 0 & i \\            
       0 & 0  & 1 & 0 \\
       0 & -1  & 0 & 0 \\
       -i & 0  & 0 &0  \end{array} & \Large{0} \end{array} \right),   
\hat{\Gamma}^7=\left( \begin{array}{cc}
           \Large{0} & \begin{array}{cccc}
                     0 & 1 & 0 & 0 \\ 
                    -1 & 0 & 0 & 0 \\
                     0 & 0 & 0 & -i \\
                     0 & 0 & i & 0  \end{array}\\
\begin{array}{cccc}
       0 & 1 & 0 & 0 \\            
       -1 & 0  & 0 & 0 \\
       0 & 0  & 0 & i \\
       0 & 0  & -i &0  \end{array} & \Large{0} \end{array} \right), \nonumber\\
\hat{\Gamma}^8=\left( \begin{array}{cc}
           \Large{0} & \begin{array}{cccc}
                     0 & 0 & 1 & 0 \\ 
                     0 & 0  & 0 & i \\
                     -1 & 0  & 0 & 0 \\
                     0 & -i  & 0 &0  \end{array}\\
\begin{array}{cccc}
       0 & 0 & 1 & 0 \\            
       0 & 0  & 0 & -i \\
      -1 & 0  & 0 & 0 \\
       0 & i  & 0 &0  \end{array} & \Large{0} \end{array} \right),   
\hat{\Gamma}^9=\left( \begin{array}{cc}
           \Large{0} & \begin{array}{cccc}
                     0 & 0 & 0 & 1 \\ 
                     0 & 0 & -i & 0 \\
                     0 & i & 0 & 0 \\
                     -1 & 0 & 0 & 0  \end{array}\\
\begin{array}{cccc}
       0 & 0 & 0 & 1 \\            
       0 & 0  & i & 0 \\
       0 & -i  & 0 & 0 \\
       -1 & 0  & 0 &0  \end{array} & \Large{0} \end{array} \right). \nonumber
\eeqn
We have checked that they in fact form the Clifford algebra.
 
  Let us turn to the remaining part $\Delta S$ of the action.
  We are interested in presenting this part in a way  $SO(2n_{f})$ flavour
 symmetry is  easily seen.  To establish this, we introduce complex $2n_{f}$
 dimensional vectors
\beqn
 {\bf Q} \equiv  \left\{ \begin{array}{ll}
       Q_{(f)}\;\;, &  f=1 \sim n_{f}  \\
     F^{-1} \tilde{Q}_{(f- n_{f})}\;\;, & f=n_{f} +1 \sim 2n_{f} \;,
   \end{array}
   \right.   \;\;\;
 {\bf Q}^{*} \equiv  \left\{ \begin{array}{ll}
       Q_{(f)}^{*} \;\;, &  f=1 \sim n_{f}  \\
      \tilde{Q}_{(f- n_{f})}^{\ast} F \;\;, & f=n_{f} +1 \sim 2n_{f}  \;.
   \end{array}
   \right.     
\eeqn
 Similarly,
\beqn
  \psivector   \equiv  \left\{ \begin{array}{ll}
      \psi_{ Q_{(f)}} \;\;, &  f=1 \sim n_{f}  \\
     F^{-1} \psi_{\tilde{Q}_{(f- n_{f})} } \;\;, & f=n_{f} +1 \sim 2n_{f} \;, 
   \end{array}
   \right.     
   \psivector^{\ast}    \equiv  \left\{ \begin{array}{ll}
       \overline{\psi}_{Q_{(f)} } \;\;, &  f=1 \sim n_{f}  \\
    \overline{\psi}_{\tilde{Q}_{(f- n_{f})} } F \;\;, & f=n_{f} +1
 \sim 2n_{f} \;.
   \end{array}
   \right.    
\eeqn
  We denote the $f$-th components of these vectors  by ${\bf Q}_{(f)}$
 etc. in the text.  After some algebras, we  find
\beqn 
   \Delta S &=& \Delta S_{b} + \Delta S_{f} =
 \left( S_{g-s} + {\cal V}_{scalar} + S_{mass} \right)
   + \left( S_{g-f} + S_{Yukawa} \right)  \;\;,  \\
 S_{g-s} &=&  -\frac{1}{g^{2}} tr \left( \sum_{\nu = 0,1,2,3,4,7}
  v_{\nu} v^{\nu} +  \sum_{I=2,3} \left[ \Phi_{I}, \Phi^{\dagger I} \right] 
  \right)   {\bf Q} \cdot {\bf Q}^{*}   \nonumber \\
     &+&  \frac{1}{g^{2}} tr \left[ \Phi_{2}, \Phi_{3} \right] F^{-1}
 {\bf Q}^{*} \cdot \Sigma  {\bf Q}^{*}
  - \frac{1}{g^{2}} tr \left[ \Phi_{2}^{\dagger}, \Phi_{3}^{\dagger}
 \right]
 {\bf Q} \cdot \Sigma F {\bf Q} \;\;,  \\
 S_{mass} &=& 
  - \frac{1}{g^{2}} tr \left( {\bf Q} \cdot M^{2} {\bf Q}^{*} \right)
  -\frac{2}{g^{2}} tr \left(v_{4} {\bf Q} \cdot M {\bf Q}^{*} \right)\;\;, \\
   {\cal V}_{scalar} &=&   - \frac{1}{2g^{2}} tr {\bf Q} \cdot
  \Sigma {\bf Q} 
  {\bf Q}^{*} \cdot  \Sigma  {\bf Q}^{*} 
 - \frac{1}{8g^{2}} tr \left[
 {\bf Q} \cdot {\bf Q}^{*} - F^{-1} {\bf Q}^{*} \cdot {\bf Q} F \right]^{2}
 \;,  \\
  S_{g-f} &=& \frac{1}{g^{2}} \left\{  \psivector^{\ast}
  \overline{\sigma}^{m} v_{m} \cdot \psivector
 + i \sqrt{2} {\bf Q}^{*} \lambda \cdot
  \psivector  - i \sqrt{2} \psivector^{\ast}
 \overline{\lambda} \cdot {\bf Q}   \right\} \;,  \\
 S_{Yukawa} &=& - \frac{1}{g^2}\left\{ \sum_{(c_{1}, c_{2})= 
   (Q, \tilde{Q}), (Q, \Phi_{1}), (\Phi_{1},\tilde{Q})}
 \frac{\partial^{2} W_{matter}}
{\partial C_{1} \partial C_{2}} \psi_{C_{2}} \psi_{C_{1}} + h.c. \right\}
   \nonumber \\
 &=& \frac{1}{g^2}\left( \frac{1}{2}   \psivector  \cdot  \Sigma   F
  \left( \sqrt{2} \Phi_{1} + M \right) \psivector
 + \sqrt{2} {\bf Q} \cdot  \Sigma  F \psi_{\Phi_{1}}  \psivector
  +  h.c. \right) \;\;.
\eeqn
  Here
\beqn
  \Sigma  &\equiv&  \left(
 \begin{array}{cc}
   0 & I  \\
   I & 0  
  \end{array}   \right)\;\;,  \\
M &\equiv&  {\rm diag} \left( m_{(1)}, \cdots, m_{(n_f)}
 - m_{(1)}, \cdots, -m_{(n_f)} \right) \;\;,  \\
  W_{{\rm matter}} &=&  \sum_{f=1}^{n_{f}} \left( m_{(f)}
 \tilde{Q}_{(f)} Q_{(f)}+ \sqrt{2} \tilde{Q}_{(f)} \Phi Q_{(f)} \right)\;\;,
\eeqn
  and $\cdot$ implies the standard inner product with respect to the
  $2n_{f}$ flavour indices.

\subsection{The kinetic terms of Schwinger-Dyson equations}
\beqn
&\bullet& \delta_{X} \Phi[(1); X^r]  \nonumber \\
&&X^r=v_{m}^{r}:  \nonumber \\
&& Tr(([v_M,[v^m,v^M]]
-\frac{1}{2} \{ \bar{\Psi}\Gamma^m, \Psi \} ) 
U[p^{(1)}_{.},\eta^{(1)}_{.};n^{(1)}_{1},n^{(1)}_{1}+1])
  \label{eq:app30}   \\ 
&&-\frac{1}{2} \Qvector^{\ast} 
U[p^{(1)}_{.},\eta{(1)}_{.},;n^{(1)}_{1},n^{(1)}_{1}+1] v^m \cdot \Qvector
+\frac{1}{2} \Qvector^{\ast}
U[\mp p^{(1)}_{.},\mp \eta;n^{(1)}_{1},n^{(1)}_{1}+1] 
v^m \cdot \Qvector \nonumber\\
&&-\frac{1}{2} \Qvector^{\ast} v^m
U[p^{(1)}_{.},\eta{(1)}_{.},;n^{(1)}_{1},n^{(1)}_{1}+1]  \cdot \Qvector
+\frac{1}{2} \Qvector^{\ast} v^m
U[\mp p^{(1)}_{.},\mp \eta;n^{(1)}_{1},n^{(1)}_{1}+1] 
\cdot \Qvector \nonumber\\
&& +\frac{1}{2} \psivector^{\ast}\cdot \bar{\sigma}^m
U[p^{(1)}_{.},\eta^{(1)}_{.};n^{(1)}_{1},n^{(1)}_{1}+1] \psivector
-\frac{1}{2}\psivector^{\ast}\cdot \bar{\sigma}^m
U[\mp p^{(1)}_{.},\mp \eta^{(1)}_{.};n^{(1)}_{1}+1,n^{(1)}_{1}] 
\psivector \;\;,  \nonumber\\
 && X^r=v_{4}^{r}: \nonumber \\
&& Tr(([v_M,[v^4,v^M]]
-\frac{1}{2} \{ \bar{\Psi}\Gamma^4, \Psi \} ) 
U[p^{(1)}_{.},\eta^{(1)}_{.};n^{(1)}_{1},n^{(1)}_{1}+1])  \\
&&-\frac{1}{2} \Qvector^{\ast} 
U[p^{(1)}_{.},\eta{(1)}_{.},;n^{(1)}_{1},n^{(1)}_{1}+1] v^m \cdot \Qvector
+\frac{1}{2} \Qvector^{\ast}
U[\mp p^{(1)}_{.},\mp \eta;n^{(1)}_{1},n^{(1)}_{1}+1] 
v^m \cdot \Qvector \nonumber\\
&&-\frac{1}{2} \Qvector^{\ast} v^m
U[p^{(1)}_{.},\eta{(1)}_{.},;n^{(1)}_{1},n^{(1)}_{1}+1]  \cdot \Qvector
+\frac{1}{2} \Qvector^{\ast} v^m
U[\mp p^{(1)}_{.},\mp \eta;n^{(1)}_{1},n^{(1)}_{1}+1] 
\cdot \Qvector \nonumber\\
&&-\Qvector^{\ast} 
U[p^{(1)}_{.},\eta{(1)}_{.},;n^{(1)}_{1},n^{(1)}_{1}+1]  \cdot M \Qvector
+\frac{1}{2} \Qvector^{\ast}
U[\mp p^{(1)}_{.},\mp \eta;n^{(1)}_{1},n^{(1)}_{1}+1] 
\cdot M \Qvector \nonumber\\
&& +\frac{1}{4} \psivector
F U[p^{(1)}_{.},\eta^{(1)}_{.};n^{(1)}_{1},n^{(1)}_{1}+1] 
\cdot \Sigma  \psivector
-\frac{1}{4}\psivector
F U[\mp p^{(1)}_{.},\mp \eta^{(1)}_{.};n^{(1)}_{1}+1,n^{(1)}_{1}] 
\cdot \Sigma F \psivector \nonumber\\
&& +\frac{1}{4} \psivector^{\ast} 
U[p^{(1)}_{.},\eta^{(1)}_{.};n^{(1)}_{1},n^{(1)}_{1}+1] F^{-1}
\cdot \Sigma  \psivector^{\ast} \nonumber \\
&& -\frac{1}{4}\psivector^{\ast}
U[\mp p^{(1)}_{.},\mp \eta^{(1)}_{.};n^{(1)}_{1}+1,n^{(1)}_{1}] F^{-1}
\cdot \Sigma  \psivector^{\ast}  \;\;, \nonumber \\
&& X^r=v_{7}^{r}: \nonumber \\
&&Tr(([v_M,[v^7,v^M]]
-\frac{1}{2} \{ \bar{\Psi}\Gamma^7, \Psi \} ) 
U[p^{(1)}_{.},\eta^{(1)}_{.};n^{(1)}_{1},n^{(1)}_{1}+1]) \\
&&-\frac{1}{2} \Qvector^{\ast} 
U[p^{(1)}_{.},\eta{(1)}_{.},;n^{(1)}_{1},n^{(1)}_{1}+1] v^m \cdot \Qvector
+\frac{1}{2} \Qvector^{\ast}
U[\mp p^{(1)}_{.},\mp \eta;n^{(1)}_{1},n^{(1)}_{1}+1] 
v^m \cdot \Qvector \nonumber\\
&&-\frac{1}{2} \Qvector^{\ast} v^m
U[p^{(1)}_{.},\eta{(1)}_{.},;n^{(1)}_{1},n^{(1)}_{1}+1]  \cdot \Qvector
+\frac{1}{2} \Qvector^{\ast} v^m
U[\mp p^{(1)}_{.},\mp \eta;n^{(1)}_{1},n^{(1)}_{1}+1] 
\cdot \Qvector \nonumber\\
&& +\frac{i}{4} \psivector
F U[p^{(1)}_{.},\eta^{(1)}_{.};n^{(1)}_{1},n^{(1)}_{1}+1] 
\cdot \Sigma  \psivector
-\frac{i}{4}\psivector
F U[\mp p^{(1)}_{.},\mp \eta^{(1)}_{.};n^{(1)}_{1}+1,n^{(1)}_{1}] 
\cdot \Sigma F \psivector \nonumber\\
&& -\frac{i}{4} \psivector^{\ast} 
U[p^{(1)}_{.},\eta^{(1)}_{.};n^{(1)}_{1},n^{(1)}_{1}+1] F^{-1}
\cdot \Sigma  \psivector^{\ast}
+\frac{i}{4}\psivector^{\ast}
U[\mp p^{(1)}_{.},\mp \eta^{(1)}_{.};n^{(1)}_{1}+1,n^{(1)}_{1}] F^{-1}
\cdot \Sigma  \psivector^{\ast} \;\;,  \nonumber\\
&& X^r=\Phi_2: \nonumber \\
&& Tr(([v_M,[\Phi_{2}^{\dagger},v^M]]
-\frac{1}{2} \{ \bar{\Psi}\frac{1}{\sqrt{2}}(\Gamma^5-i\Gamma^6), \Psi \} ) 
U[p^{(1)}_{.},\eta^{(1)}_{.};n^{(1)}_{1},n^{(1)}_{1}+1]) \\
&&+\frac{1}{2} \Qvector^{\ast} 
U[p^{(1)}_{.},\eta{(1)}_{.},;n^{(1)}_{1},n^{(1)}_{1}+1] \Phi_3 F^{-1}
\cdot \Sigma \Qvector^{\ast}
+\frac{1}{2} \Qvector^{\ast}
U[\mp p^{(1)}_{.},\mp \eta;n^{(1)}_{1},n^{(1)}_{1}+1] \Phi_3 F^{-1}
\cdot \Sigma \Qvector^{\ast} \nonumber\\
&& -\frac{1}{2} \Qvector^{\ast} \Phi_3
U[p^{(1)}_{.},\eta{(1)}_{.},;n^{(1)}_{1},n^{(1)}_{1}+1]  F^{-1}
\cdot \Sigma \Qvector^{\ast}
-\frac{1}{2} \Qvector^{\ast} \Phi_3
U[\mp p^{(1)}_{.},\mp \eta;n^{(1)}_{1},n^{(1)}_{1}+1]  F^{-1}
\cdot \Sigma \Qvector^{\ast}\;\;,  \nonumber \\
&& X^r=\Phi_3: \nonumber \\
&&Tr(([v_M,[\Phi_{3}^{\dagger},v^M]]
-\frac{1}{2} \{ \bar{\Psi}\frac{1}{\sqrt{2}}(\Gamma^8-i\Gamma^9), \Psi \} ) 
U[p^{(1)}_{.},\eta^{(1)}_{.};n^{(1)}_{1},n^{(1)}_{1}+1]) \\
&&-\frac{1}{2} \Qvector^{\ast} 
U[p^{(1)}_{.},\eta{(1)}_{.},;n^{(1)}_{1},n^{(1)}_{1}+1] \Phi_2 F^{-1}
\cdot \Sigma \Qvector^{\ast}
-\frac{1}{2} \Qvector^{\ast}
U[\mp p^{(1)}_{.},\mp \eta;n^{(1)}_{1},n^{(1)}_{1}+1] \Phi_2 F^{-1}
\cdot \Sigma \Qvector^{\ast} \nonumber\\
&& +\frac{1}{2} \Qvector^{\ast} \Phi_2
U[p^{(1)}_{.},\eta{(1)}_{.},;n^{(1)}_{1},n^{(1)}_{1}+1]  F^{-1}
\cdot \Sigma \Qvector^{\ast}
+\frac{1}{2} \Qvector^{\ast} \Phi_2
U[\mp p^{(1)}_{.},\mp \eta;n^{(1)}_{1},n^{(1)}_{1}+1]  F^{-1}
\cdot \Sigma \Qvector^{\ast}  \;\;, \nonumber \\
&& X^r=\lambda: \nonumber \\
&& Tr((\sigma^m [v_m,\bar{\lambda}] 
-i \sqrt{2}[\psi_I,\Phi^{\dagger}_{I}])
U[p^{(1)}_{.},\eta^{(1)}_{.};n^{(1)}_{1},n^{(1)}_{1}+1]) \\
&&  +\frac{i}{\sqrt{2}} \Qvector^{\ast} 
U[p^{(1)}_{.},\eta{(1)}_{.},;n^{(1)}_{1},n^{(1)}_{1}+1] \cdot \psivector
-\frac{i}{\sqrt{2}} \Qvector^{\ast}
U[\mp p^{(1)}_{.},\mp \eta;n^{(1)}_{1},n^{(1)}_{1}+1] 
\cdot \psivector \;\;,  \nonumber \\
&&  X^r=\psi_1:  \nonumber  \\
&& Tr((\sigma^m [v_m,\bar{\psi}^{1}] 
-i \sqrt{2}[\lambda,\Phi^{\dagger}_{1}])
U[p^{(1)}_{.},\eta^{(1)}_{.};n^{(1)}_{1},n^{(1)}_{1}+1])  \\
&& +\frac{1}{\sqrt{2}} \Qvector F 
U[p^{(1)}_{.},\eta{(1)}_{.},;n^{(1)}_{1},n^{(1)}_{1}+1] 
\cdot \Sigma \psivector
-\frac{1}{\sqrt{2}} \Qvector F
U[\mp p^{(1)}_{.},\mp \eta;n^{(1)}_{1},n^{(1)}_{1}+1] 
\cdot \Sigma \psivector  \;\;, \nonumber \\
&&X^r=\psi_I \; (I=2,3): \nonumber  \\
&& Tr((\sigma^m [v_m,\bar{\psi}^{I}] 
-i \sqrt{2}[\lambda,\Phi^{\dagger}_{I}])
U[p^{(1)}_{.},\eta^{(1)}_{.};n^{(1)}_{1},n^{(1)}_{1}+1]) \;\;.
\eeqn

\beqn
&\bullet& \delta_{X} \Psi[(1); X^r] \nonumber \\
&& X^r=v_{m}^{r}:   \nonumber \\
&& \left( \Lambda^{(1)'} \Pivector_{(f^{(1)'})} \right) F
U[k^{(1)}_{.},\zeta^{(1)}_{.};l^{(1)}_{2},l^{(1)}_{1}+1] \left( [v_M,[v^m,v^M]]
-\frac{1}{2} \{ \bar{\Psi}\Gamma^m, \Psi \} \right)  \nonumber \\
&&  U[k^{(1)}_{.},\zeta^{(1)}_{.};l^{(1)}_{1},l^{(1)}_{0}]
  \left( \Lambda^{(1)} \Pivector_{(f^{(1)})} \right) \nonumber\\
&&-\frac{1}{2}
  \left( \Lambda^{(1)'} \Pivector_{(f^{(1)'})} \right) F
U[k^{(1)}_{.},\zeta^{(1)}_{.};l^{(1)}_{2},l^{(1)}_{1}+1] 
v^m \Qvector \cdot \Qvector^{\ast}
U[k^{(1)}_{.},\zeta^{(1)}_{.};l^{(1)}_{1},l^{(1)}_{0}]
\left(  \Lambda^{(1)} \Pivector_{(f^{(1)})}  \right) \\
&& -\frac{1}{2}
\left(  \Lambda^{(1)'}  \Pivector_{(f^{(1)'})} \right) F
U[k^{(1)}_{.},\zeta^{(1)}_{.};l^{(1)}_{2},l^{(1)}_{1}+1] F^{-1} 
\Qvector^{\ast} \cdot \Qvector F v^m
U[k^{(1)}_{.},\zeta^{(1)}_{.};l^{(1)}_{1},l^{(1)}_{0}]
 \left( \Lambda^{(1)} \Pivector_{(f^{(1)})} \right)  \nonumber \\
&&-\frac{1}{2}
  \left( \Lambda^{(1)'} \Pivector_{(f^{(1)'})} \right) F
U[k^{(1)}_{.},\zeta^{(1)}_{.};l^{(1)}_{2},l^{(1)}_{1}+1] 
\Qvector \cdot \Qvector^{\ast} v^m
U[k^{(1)}_{.},\zeta^{(1)}_{.};l^{(1)}_{1},l^{(1)}_{0}]
\left(  \Lambda^{(1)} \Pivector_{(f^{(1)})} \right)  \nonumber\\
&&-\frac{1}{2}
\left(  \Lambda^{(1)'} \Pivector_{(f^{(1)'})} \right) F
U[k^{(1)}_{.},\zeta^{(1)}_{.};l^{(1)}_{2},l^{(1)}_{1}+1]  v^m F^{-1} 
\Qvector^{\ast} \cdot \Qvector F 
U[k^{(1)}_{.},\zeta^{(1)}_{.};l^{(1)}_{1},l^{(1)}_{0}]
 \left(  \Lambda^{(1)} \Pivector_{(f^{(1)})} \right) \nonumber\\
&&-\frac{1}{2}
 \left( \Lambda^{(1)'} \Pivector_{(f^{(1)'})} \right) F
U[k^{(1)}_{.},\zeta^{(1)}_{.};l^{(1)}_{2},l^{(1)}_{1}+1] 
\psivector \cdot \sigma^m \psivector^{\ast}
U[k^{(1)}_{.},\zeta^{(1)}_{.};l^{(1)}_{1},l^{(1)}_{0}]
\left(  \Lambda^{(1)} \Pivector_{(f^{(1)})} \right) \nonumber\\
&&-\frac{1}{2}
 \left( \Lambda^{(1)'}  \Pivector_{(f^{(1)'})} \right) F
U[k^{(1)}_{.},\zeta^{(1)}_{.};l^{(1)}_{2},l^{(1)}_{1}+1] F^{-1} 
\psivector^{\ast} \cdot \bar{\sigma}^m \psivector F
U[k^{(1)}_{.},\zeta^{(1)}_{.};l^{(1)}_{1},l^{(1)}_{0}]
 \left( \Lambda^{(1)} \Pivector_{(f^{(1)})} \right) \;\;. \nonumber
\eeqn
\beqn
&\bullet& \delta_{\Zvector} \Psi[(1);\Zvector] \nonumber \\
&&\Zvector_{(f)i}=\Qvector_{(f)i} :    \nonumber \\
&&  \left( \Qvector^{\ast}_{(f)}  \left( v_{\nu}v^{\nu}+[\Phi_I,\Phi^I] \right)
+ \left(\Qvector \Sigma \right)_{(f)}F[\Phi_{2}^{\dagger},\Phi_{3}^{\dagger}]
+ \left( \Qvector^{\ast}M^2 \right)_{(f)}+2 \left( \Qvector^{\ast}M
 \right)_{(f)}v_4
- i \sqrt{2} \psivector^{\ast}_{(f)}\bar{\lambda}  \right.   \nonumber\\
&& \left.  -\sqrt{2} \left( \psivector \Sigma \right)_{(f)} F \psi_{\Phi_1} 
+ \left( \Qvector \Sigma \right)_{(f)} \Qvector^{\ast} \cdot \Sigma
 \Qvector^{\ast}
+\frac{1}{2} \left( \Qvector^{\ast}_{(f)} \Qvector \cdot \Qvector^{\ast}
-\Qvector^{\ast}_{(f)} F^{-1} \Qvector^{\ast} \cdot \Qvector F \right)  \right)
  \\
&& U[k^{(1)}_{.},\zeta^{(1)}_{.};l^{(1)}_{1},l^{(1)}_{0}]
 \left( \Lambda^{(1)} \Pivector_{f^{(1)}} \right)
 \;\;, \nonumber \\
&&\Zvector_{(f)i}=\psivector_{(f)i \alpha}:  \nonumber \\
&&  \left(  \psivector^{\ast}_{(f)} \bar{\sigma}^{m} v_{m}
+i \sqrt{2} \Qvector_{(f)}^{\ast} \lambda  \right.  \\
&& \left. + \left( \psivector  \Sigma F \left( \sqrt{2}\Phi_1 + M \right)
 \right)_{(f)}
+\sqrt{2} \left( \Qvector  \Sigma F \psi_{\Phi_1} \right)_{(f)}  \right)
U[k^{(1)}_{.},\zeta^{(1)}_{.};l^{(1)}_{1},l^{(1)}_{0}] 
 \left( \Lambda^{(1)} \Pivector_{f^{(1)}} \right) \;\;. \nonumber
\eeqn

\end{document}